\documentclass[12pt]{article}
\usepackage{graphicx}
\usepackage{rotating}
\usepackage{amsmath}
\usepackage[a4paper]{geometry}
\usepackage{setspace}

\begin{document}

\title{The Structure and Growth of Weighted Networks}

\author{Massimo Riccaboni\thanks{Department of Computer and Management Sciences, 
University of Trento (Italy) and Department of Physics, Boston University (United States). E-mail: massimo.riccaboni@unitn.it} 
\and Stefano Schiavo\thanks{Department of Economics, University of Trento (Italy) and 
OFCE (France). E-mail: stefano.schiavo@unitn.it}\thanks{We acknowledge Gene Stanley, Sergey Buldyrev, Fabio Pammolli, Jakub Growiec, 
Dongfeng Fu, Kazuko Yamasaki, Kaushik Matia, Lidia Ponta, Giorgio Fagiolo and Javier Reyes for previous work 
on which this contribution builds.}}
\date{This version --- December 2009}
\maketitle

\vspace{-12pt}
\begin{abstract}
\noindent 

\noindent We develop a simple theoretical framework for the evolution of weighted networks that is consistent with a number of stylized features of real-world data. In our framework, the Barab\'{a}si-Albert model of network evolution is extended by assuming that link weights evolve according to a geometric Brownian motion. 
Our model is verified by means of simulations and real world trade data. 
We show that the model correctly predicts the intensity and growth distribution of links, 
the size-variance relationships of the growth of link weights, 
the relationship between the degree and strength of nodes, as well as the scale-free structure of the network.

\vspace{6pt}


\end{abstract}
\thispagestyle{empty}


\newpage
\section{Introduction}
\label{intro}

Graph theory has been used to describe a vast array of real-world phenomena, but only recently the attention has shifted from binary to weighted  graphs, from both an empirical \cite{newm01,pave04,guimc05,garlc05,fagic08,bhatt08} and a theoretical perspective \cite{yook01,zheng03,barrc04}.
The empirical literature has come up with a number of robust stylized facts that apply to a wide range of phenomena as different as Internet traffic, airport connections, and international trade. 
In particular, it has been demonstrated that weighted graphs display 
(i) a power law connectivity distribution $P(K)$, with finite size truncation \cite{bbpsv04,guimc05}; 
(ii) a skewed distribution of link weights $P(w)$ and node strengths measured as the sum of the weights of the links of a given node $P(W)$ \cite{rago07,sbv09}; 
(iii) a power-law relation between node strength $W$ and node degree $K$: $W=K^\theta$, with $\theta$ ranging between 1.3 and 1.5 \cite{bbpsv04,eom08}.  

In this paper we present a simple stochastic model of proportionate growth of both the number and the weight of links to describe the structure and evolution of weighted networks and account for above mentioned regularities.
In our setup we extend the Barab\'{a}si and Albert (BA) model \cite{baal99} to accommodate {\em weighted} network dynamics.
This is done exploiting the theoretical framework recently put forward by Stanley and co-authors to explain the scaling distribution of fluctuations in complex systems \cite{Kazuko,fuc05,ricc08pnas}. 

We test our model using data on the network of international trade flows, which  is a prototypical example of a real-world network that is inherently weighted.
International trade flows have been traditionally analyzed in the contest of the so-called gravity model \cite{tinb62} that relates bilateral flows to countries' size and distance. However, one of the main limits of this approach is its inability to capture the large fraction of zeros existing in the matrix of bilateral links. 
Although this has recently been addressed in the context of standard economic theory \cite{hmr08}, graph theory as been applied to naturally accommodate this feature of the data.

We selected the international trade network (ITN) as a testbed for our model based on the following considerations. 
First, the ITN has been already extensively investigated \cite{fagic08,bhatt08,SeBo03,Garla2004,Garla2005,Bhatta2007a,fagiC09pre}, and previous works on the ITN provide us with a rich set of empirical regularities. 
Thus, we know that the link weight distribution assumes a log-normal form in the case of the ITN \cite{Bhatta2007a,fagiC09pre}, whereas their growth rates display fat tails \cite{fagiC09pre}. 
Second, the relationship between node strength and degree is crucial in the economic literature about ITN since it is related to the interplay between intensive and extensive margins of trade, which is key to explain  trade flows \cite{Chaney08}.\footnote{The extensive margin consists of the number of trading partners and the number of products exported $K$, whereas the intensive margin represents the amount shipped per product per country $w$.} 
Third, despite the structural inertia of the ITN, the huge volatility of trade flows after the 2008 global financial crisis has recently attracted a great deal of attention. Our theoretical framework provides an explanation for the relationship 
between node centrality and the variance of network flows.

The paper is organized as follows. 
Section \ref{sec:model} presents the model and its most important predictions. 
We then test our model using data on the ITN (Section \ref{sec:emp}) and simulations (Section \ref{sec:sim}).  
Finally, in the last Section we lie down some conclusions and outline possible patterns for future research.

\section{The model}
\label{sec:model}

Barab\'{a}si and Albert \cite{baal99} have proposed a simple stochastic model of network growth based on
 preferential attachment  which accounts for many of the stylized facts observed in real-world networks.
The increasing interest in the study of \emph{weighted} versions of networks calls for an extension of the original BA model to account for the large degree of heterogeneity across link weights \cite{yook01,barrc04}. 
The route we take here exploits the theoretical framework recently put forward by Stanley and co-authors \cite{fuc05}  to deal with the growth dynamics of complex systems. 
We prove that our model is capable to accurately match the structural properties that characterize a number of real-world weighted networks. 

We therefore propose a generalized version of the BA model to describe the dynamic and growth of weighted networks, by modeling them as a set of links of different weight occurring among nodes. 
In particular, we assume that the weight of links grows according to a geometric Brownian motion 
(also known as Gibrat's \emph{law of proportionate effects} \cite{gibr31}), so that the expected value of 
the growth rate of link weights is independent of their current level.

The key sets of assumptions in the model are the following \cite{baal99,bori03,fuc05}:
\begin{enumerate}
        \item[1.] The network begins at time $t=0$ with $N_0$ nodes each with a self loop. 
At each time step $t= \left\{ {1,\dots,M} \right\}$, a new link among two nodes arises: thus the number of links (excluding self-loops which are used only for initialization) existing at time $t$ is $m_t = t$. 
We write $K_i(t)$ for the number of links of node $i$ at time $t$ (node degree). 
To identify the nodes connected by the newly formed link at time $t$ we adopt the following procedure: with probability $a$ the new link is assigned to a new source node, whereas with probability $1-a$ it is allocated to an existing node $i$. 
In the latter case, the probability of choosing node $i$ is given by: $p_i(t)=K_i(t-1)/2t$.
Edge endpoints $i$ and $j$ of the new link are chosen symmetrically with $i\neq j$. 
Thus with probability $a$ the new link is assigned to a new target node, while with probability $1-a$ it is allocated to an existing node with probability $p_j(t)=K_j(t-1)/(2t-K_i(t-1))$ if $j\neq i$ and $p_j(t)=0$ otherwise. 
Hence, at each time $t$ this rule identifies the pair of (distinct) nodes to be linked;

      \item[2.] at time $t$ each existing link between nodes $i$ and $j$ has weight $w_{ij}(t)>0$,        where $K_i$, $K_j$ and $w_{ij}$ are independent random variables. 
At time $t+1$ the weight of each link is increased or decreased by a random factor $x_{ij}(t)$, so that $w_{ij}(t+1)= w_{ij}(t)x_{ij}(t)$. 
The shocks  and initial link weights are taken from a distribution with finite mean and standard deviation.
        \end{enumerate}

\noindent Thus we assume that each link weight grows in time according to a random process. 
Moreover the two processes governing link formation and weight growth are assumed to be independent.
We therefore combine a preferential attachment mechanism (Assumption 1), with an independent geometric Brownian motion of link weights (Assumption 2).
In this way we obtain a generalization of the BA setup capable to account for the growth of weighted networks.

Based on the first assumption we derive the degree distribution $P(K)$ \cite{baal99,ricc07jeea}. 
In the absence of the entry of new nodes ($a=0$) the probability distribution of the number of links at
large $t$, i.e. the distribution $P(K)$, is exponential:
\begin{equation}
P(K)\approx \frac{1}{\bar K}\exp(-K/\bar K),
\label{P_K_old}
\end{equation}
where $\bar K=2t/N_0$ is the average number of links per node, which linearly grows with time.\footnote{$\bar K$ does not include initial self loops.}

If $a>0$, $P(K)$ becomes a Yule distribution which behaves as a power law for small $K$:
\begin{equation}
P(K)\sim K^{-\varphi},
\end{equation}
where $\varphi=2+a/(1-a)\ge 2$, followed by the exponential decay of Eq.(\ref{P_K_old}) 
for large $K$ with $\bar K=(1+2t/N_0)^{1-a}-1$ \cite{Kazuko}.  

Hence, in the limit of large $t$ when $a=0$ (no entry), the distribution of $P(K)$ converges to an exponential; on the contrary when $a>0$ and small the connectivity distribution at large $t$ converges to a power-law with an exponential cutoff \cite{Kazuko}.

By using the second assumption we can compute the growth rate of the strength of nodes.
The strength of node $i$ is given by $W_i= \sum\limits_{K_i}{w_{ij}}$. 
The growth rate is measured as $g=\ln(W(t+1)/W(t))$. 
Thus, the resulting distribution of the growth rates of node strength $P(g)$ is determined by
 \begin{equation}
 P(g) \equiv \sum_{K=1}^{\infty}P(K)P(g|K),
 \label{P_g_g_sum}
 \end{equation}
 where $P(K)$ is the connectivity distribution,
 computed in the previous stage of the model and $P(g|K)$ is the conditional
 distribution of growth rates of nodes with given number of links determined
 by the distribution $P(w)$ and $P(x)$.

Fu and colleagues \cite{fuc05} find an analytical solution  for the distribution of the growth rates of the weights of links $P(g)$  for the case when  $a \rightarrow 0$ and $t\to\infty$, 

\begin{equation}
P(g) \approx \frac{2V_g}{\sqrt{g^2+2V_g}\,(|g|+\sqrt{g^2+2V_g})^2}
\label{p_new_1}
\end{equation}

$P(g)$ has similar behavior to the Laplace distribution for
small $g$ i.e. $P(g)\approx \exp(-\sqrt{2}|g|/\sqrt{V_g})/\sqrt{2V_g}$,
while for large $g$, $P(g)$ has power law tails $P(g)\sim g^{-3}$ which
are eventually truncated for $g\to \infty$ by the distribution $P(x)$
of the growth rate of a single link.

A further implication of the model that can be derived from the second assumption concerns the distribution of  link weights $P(w)$.
The proportional growth process (Assumption 2) implies that the distribution of the weights $P(w)$ converges 
to a log-normal. 
Thus  node strength $W$ is given by the sum of $K$ log-normally distributed stochastic values. 
Since the log-normal distribution is not stable upon aggregation, the distribution of node strength $P(W)$ is multiplied by a stretching factor that, depending on the distribution of the number of links $P(K)$ could lead to a Pareto upper tail \cite{riccel}.

Moreover, a negative relationship exits among the weight of links and the variance of their growth rate. 
Our model implies an approximate power-law behavior for the variance of growth rates of the form $\sigma(g)=W^{-\beta(W)}$ where $\beta(W)$ is an exponent that weakly depends on the strength $W$. 
In particular, $\beta=0$ for small values of $W$, $\beta=1/2$ for  $W \rightarrow \infty$, and it is 
well approximated by $\beta \approx 0.2$ for a wide range of intermediate values of $W$ \cite{ricc08pnas} .

Finally, the model yields a prediction also on the relation between the degree $K$ and the strength $W$ of each node.
In Section \ref{sec:sim} we show that since the weight of each link is sampled from a log-normal distribution ($w$ are log-normally distributed), and given the skewness of such a density function, the law of large numbers does not work effectively. 
In other words, the probability to draw a large value for a link weight increases with the number of draws, 
thus generating a positive power law relationship between $W$ and $K$, for small $K$.

\section{The Empirical Evidence}
\label{sec:emp}

To test our model we use the NBER-United Nations Trade Data \cite{feenc05} available through the 
Center for International Data at UC Davis. This database provides bilateral trade flows among  countries over 1962--2000, disaggregated at the level of commodity groups (4-digit level of the
Standard International Trade Classification, SITC). Data are in thousands US dollars and, for product-level flows, there is a lower threshold at 100,000 dollars below which transactions are not recorded.
One point to note is that disaggregated data are not always consistent with country trade flows: in a number of cases we do not observe any 4-digit transaction recorded between two countries, but nevertheless find a positive total trade, and vice-versa.
Since we take the number of product traded among any two country pairs as the empirical counterpart of the 
number of transactions, to avoid inconsistency we compute the total trade by aggregating commodity-level data.


In this Section we test the predictions of our model while in the following Section, we use the data to calibrate the simulations and check for the ability of the model to  replicate real-world phenomena by comparing simulated and actual trade flows.
We already know from previous work \cite{fagiC09pre} that the main features of the ITN are broadly consistent with our model.
Here we look in more details at some specific characteristics of ITN.

\begin{figure}[htbf]
\begin{center}
 \includegraphics[width = 0.9 \textwidth]{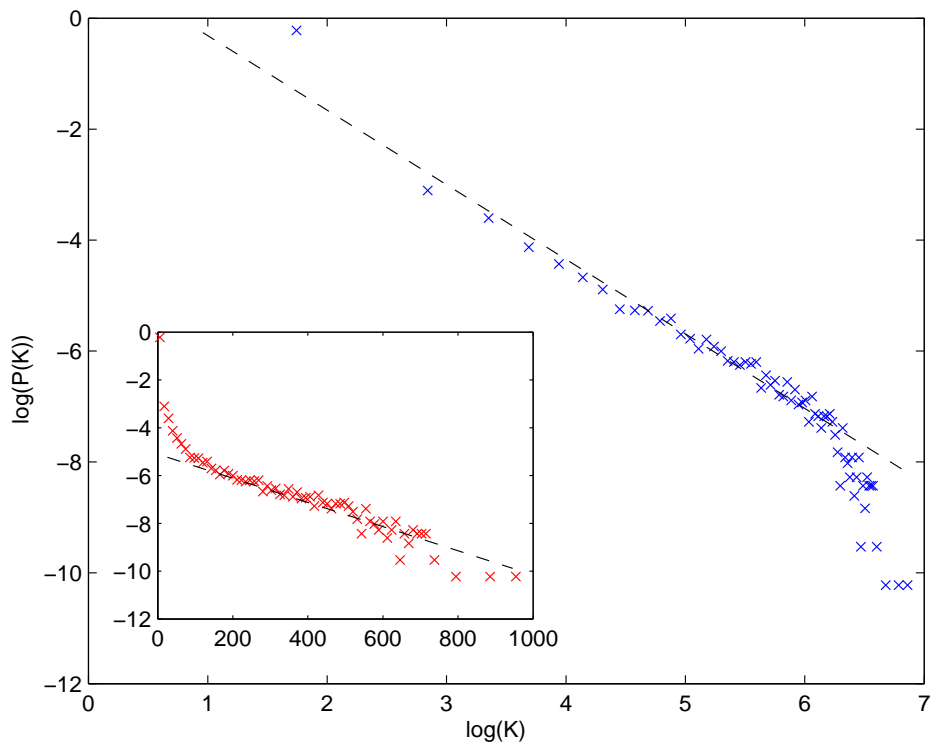}
 \caption{Distribution of the number of products traded -- 1997. Double logarithmic scale (main plot) and semi logarithmic scale (inset)}
 \label{fig:k_distr}
\end{center}
\end{figure}

Figure \ref{fig:k_distr} shows that the distribution $P(K)$, that is the number of 4-digit SITC products traded by countries, is power-law distributed with an exponential cutoff.
The main plot displays the probability distribution in log-log scale, whereby the power-law is the straight 
line body, and the exponential cutoff is represented by the right tail.
The inset presents the same distribution in semi-log scale: this time it is the exponential part of the distribution that becomes a straight line, so that with this trick we can magnify what happens to the probability distribution as $K$ grows large.
As discussed in Section \ref{sec:model} above, the power law distribution of $K$ hints at the existence of moderate entry of new nodes in the network. Indeed, 17 new countries enter into the ITN during the observed time frame, mostly due to the collapse of the Soviet Union and Yugoslavia.   

\begin{figure}[bhft]
\begin{center}
\includegraphics[width = 0.9 \textwidth]{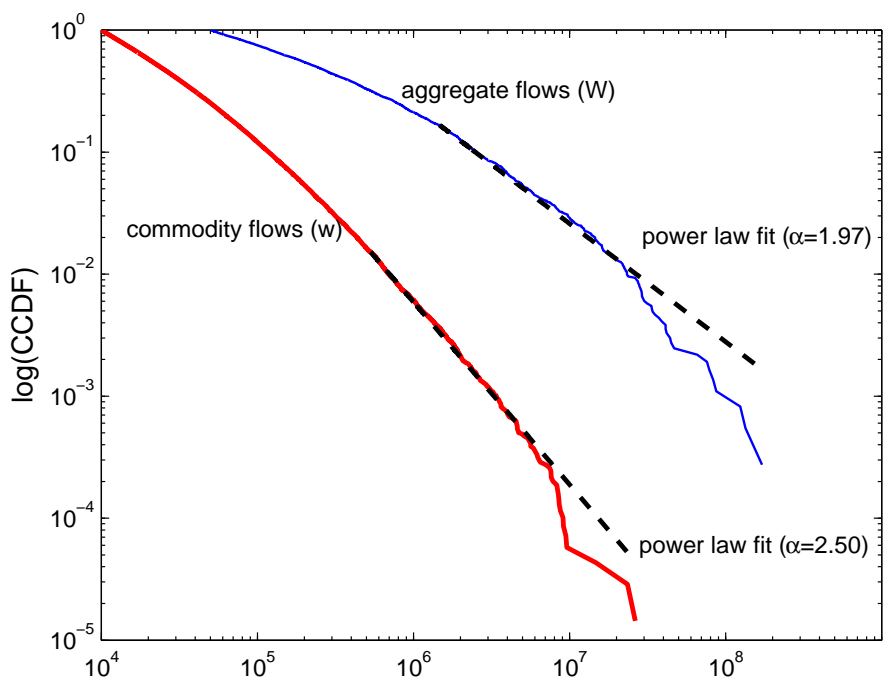}
\caption{The distribution of the link weights and node strength in year 1997. Complementary cumulative distribution of the strength distribution $P(W)$ (aggregate flows) and link weights $P(w)$ (commodity flows) and their power-law fits (dashed lines) \cite{clauc09}}
\label{fig:w_distr}
\end{center}
\end{figure}

Moving to the weighted version of the network, one can look at the distribution of positive link weights as measured by bilateral trade flows at the commodity level, $P(w)$, as well as the total value of country trade or node strength $P(W)$.
Figure \ref{fig:w_distr} shows the complementary cumulative probability distribution of trade flows in log-log scale, both for product-level transactions and for aggregate flows.
Figure \ref{fig:w_distr} refers to 1997 data (other years display the same behavior).
We observe that both distributions show the parabolic shape typical of the log-normal distribution, thus conforming to previous findings \cite{Bhatta2007a,fagiC09pre}.
As predicted, upon aggregation the power-law behavior of the upper tail become more pronounced \cite{riccel}. However,  this departure from log-normality concerns a very small number of observations (0.16\% in the case of commodities flows, 2.21\% for aggregate flows) since only a few new nodes (countries) enter the network over time.

\begin{figure}[tbhf]
\begin{center}
\includegraphics[width = 0.9 \textwidth]{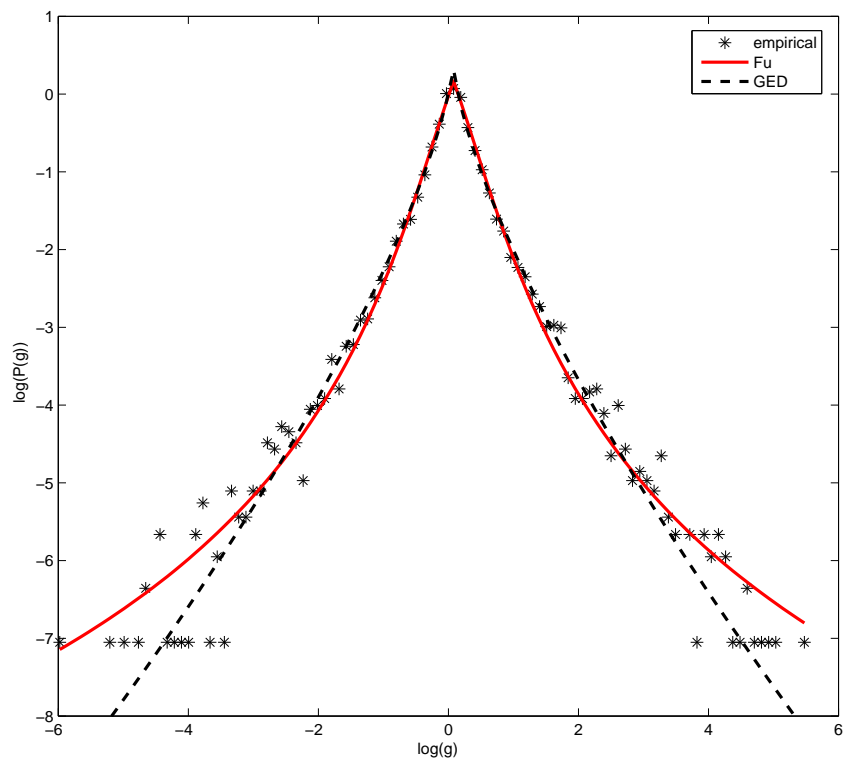}
\caption{Distribution of the growth rates of aggregate trade flows $P(g)$}
\label{fig:g_macro}
\end{center}
\end{figure}

As for the growth of trade flows, Figure \ref{fig:g_macro} shows  the empirical distribution $P(g)$ together with the maximum likelihood fit of Eq.(\ref{p_new_1}) as well as the Generalized Exponential Distribution (GED, with shape parameter 0.7224).

Goodness of fit tests, reported in Table \ref{tab:gof1}, show that $P(g)$ is neither Gaussian nor Laplace, 
whereas the distribution in Equation (\ref{p_new_1}) performs much better in terms of Kolmogorov-Smirnov (KS) and Anderson-Darling (AD) tests.\footnote{The KS and the AD are non-parametric tests used to evaluate whether a sample comes from a population with a specific distribution. Both KS and AD tests quantify a distance between the empirical distribution function of the sample and the cumulative distribution function of the reference distribution. The AD test gives more weight to the tails than the KS test. More detailed information are available in \cite{chak67,step74}.}
Hence, the growth of nodes centrality, as measured by strength $W$, follows the same law of the fluctuations of the size of complex systems \cite{fuc05,fnr08}. This is not surprising, since the size of an airport can be measured by the number of the passengers that travel through it and the size of a firm in terms of sales is given by the sum of the value of each product it sells. 
Thus the theoretical framework of Stanley and colleagues \cite{fuc05} complements and completes the BA proportional growth model in the case of weighted networks.

\begin{table}[t!]
\caption{Kolmogorov-Smirnov (KS) and Anderson-Darling (AD) goodness of fit tests for the distribution
of growth rates of trade flows $P(g)$}
\label{tab:gof1}
\centering
\begin{tabular}{lrrrrrr}
& & & & \\
\hline
    Distribution & Mean  & Variance & KS & AD \\
\hline
    Gauss     & 0.0333 & 0.8417 & 10.8305 & 117329 \\
    Laplace   & 0.0040 & 0.5338 & 2.8414 & 1.4107 \\
    GED (shape parameter 0.72) & 0.0444 & 0.2899 &  1.0915 & 0.0314 \\
    Equation (\ref{p_new_1})    & 0.0651 & 0.3658 & 0.8214 & 0.0477 \\
\hline
\end{tabular}
\end{table}

\begin{figure}[hbthf]
\centering
\includegraphics[width=14cm]{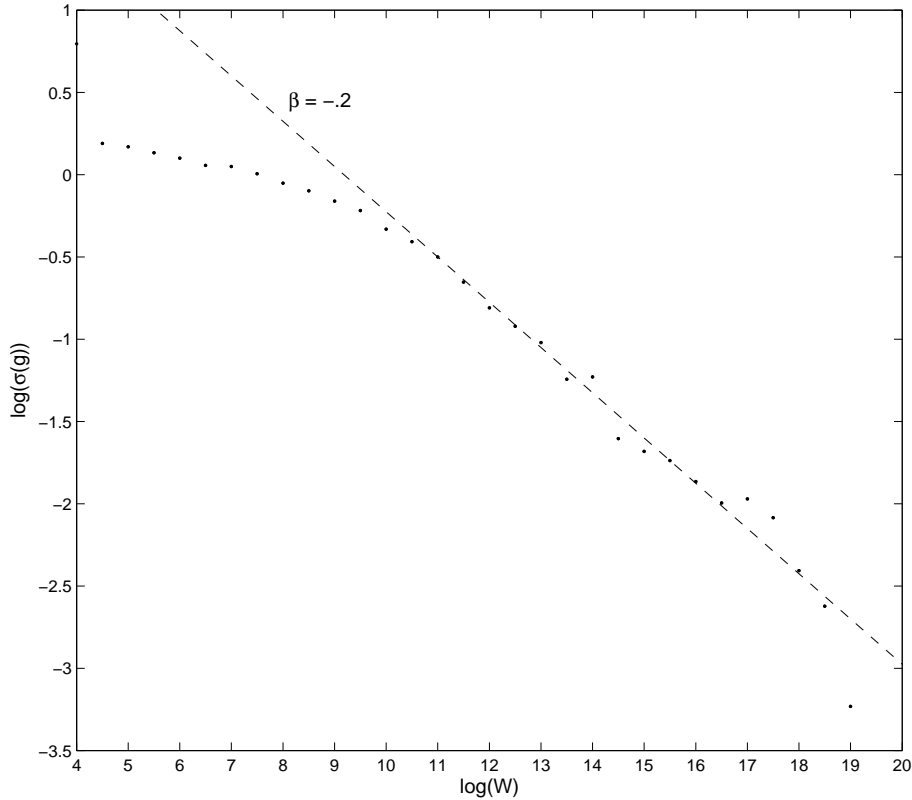}
\caption{Size-variance relationship between nodes strength $W$ (trade values) and the standard deviation of its growth rate $\sigma(g)$, double logarithmic scale}
\label{fig:size_var}
\end{figure}

As discussed in Section \ref{sec:model}, our model implies a negative relationship between node strength and the variance of its growth rate. 
Figure \ref{fig:size_var} reports the standard deviation of the annual growth rates of node strength ($g$) and their initial magnitude ($W$).
The standard deviation of the growth rate of link weights exhibits a power law relationship $\sigma(g)=W^{-\beta}$ with $\beta\approx .2$, as predicted by the  model \cite{ricc08pnas}. 
This implies that the fluctuations of the most intense trade relationships are more volatile than expected based on the central limit theorem.

All in all, our model accurately predicts the growth and weight distribution of trade flows, the number of commodities traded and the size-variance relationship of trade flows. 
Thus we can conclude that a stochastic model that assumes a proportional growth of the number of links combined with an independent proportional growth process of link weights can reproduce most of the observed structural features of the world trade web and should be taken as a valid stochastic benchmark to test the explanatory power of alternative theories of the evolution of international trade and weighted networks in general. 
In the next section we will compare the structure of random networks generated according to our model and with the real world trade network.

\section{Simulation Results}
\label{sec:sim}

Based on the assumptions in Section \ref{sec:model} we generate a set of random networks and fit them
with real world data in order to test the predictive capability of our theoretical framework. 
We will proceed in two steps. First, we generate the unweighted network according to the first set of assumptions.
Next, we assign the value of weights based on a random sampling of $K$ values from a log-normal distribution $P(w)$ whose parameters are obtained through a maximum likelihood fit of the real world data.

We model a system where at every time $t$ a new link is added, which represents the possibility to exchange one product with a trading partner. 
We slightly modify the original setting in order to account for the possibility that the new links could be assigned randomly rather than proportionally to node connectivity. 
Thus in our simulations the parameter $a$ governs the entry of new nodes according to Assumption 1, 
whereas parameter $b$ is the probability that a new link is assigned randomly. 
Thus with probability  $a$ the new link is assigned to a new source node, whereas with probability $1-a$ it is allocated to an existing node $i$. 
In the latter case, the probability of choosing node $i$ is now given by $p_i(t)=(1-b)K_i(t-1)/2t+b/N_{t-1}$ where $N_{t-1}$ is the number of nodes at time $t-1$. The target of the new link is chosen symmetrically with $i\neq j$. 

By tuning the two model parameters $a$ and $b$ we generate different networks in terms of the connectivity distribution of trade links $P(K)$. 
In particular, without entry ($a=0$) and completely random allocation of opportunities ($b=1$) one obtains a random graph characterized by a Poisson connectivity distribution \cite{erre59}, whereas allowing entry ($a>0$)  $P(K)$ is exponentially distributed. 
Keeping a positive entry rate, but assigning opportunities according to a preferential attachment model ($b=0$) the model leads to a power-law connectivity distribution with an exponential cut-off which is more pronounced the higher is the number of initial nodes $N_0$. 
In the limit case in which entry of new nodes is ruled out ($a=0$) then the connectivity distribution tends toward a  Bose-Einstein geometric distribution.

We compare the structure of random scale-free model networks with the real world trade network in 1997. 
Since the structure of  the network is highly stable over time results do not change substantially if we  compare simulations with the structure of the real world network in different years. 
In the first stage, we generate one million networks with $a$ and $b$ both ranging from 0 to 1.
We simulate random networks of 166 nodes (countries) and 1,079,398 links (number of different commodities traded by two countries). The number of commodities traded is taken as a proxy of the number of transactions. Next we select the random networks that better fit the real world pattern in terms of correlation, as measured by the Mantel $r$ test, and connectivity distribution.\footnote{The Mantel test is a non-parametric statistical test of the correlation between two matrices \cite{mant67}. The test is based on the distance or dissimilarity matrices which, in the present case, summarize the number of links between two nodes in the simulated and real networks. A typical use of the test entails comparing an observed  connectivity matrix with one posed by a model. The significance of a correlation is evaluated via permutations, whereby the rows and columns of the matrices are randomly rearranged.}

\begin{figure}[hbtf]
\begin{center}
 \includegraphics[width = 0.9 \textwidth]{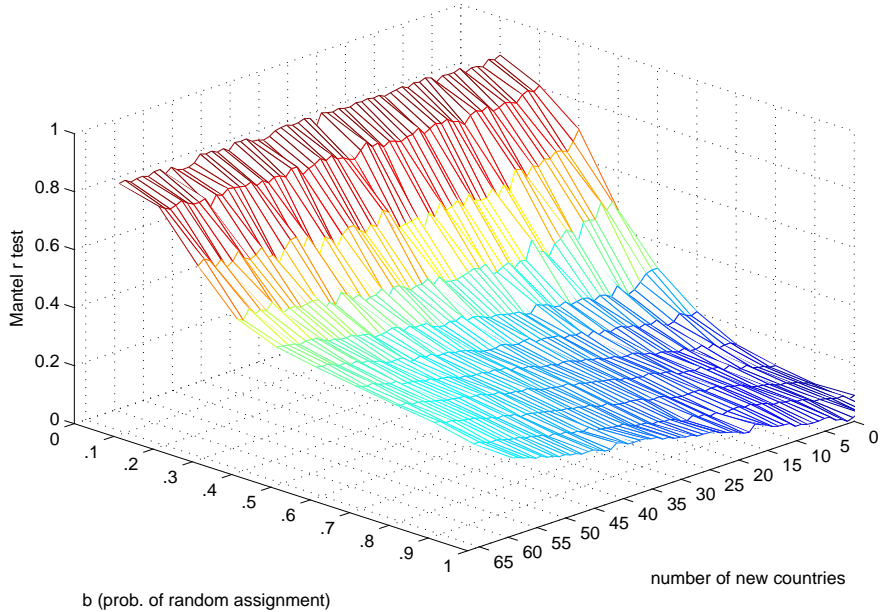}
 \caption{Mantel test comparing simulated and real networks}
 \label{fig:mantel}
\end{center}
\end{figure}

\begin{figure}[tbhf]
\centering
\includegraphics[width=14cm,height=9cm]{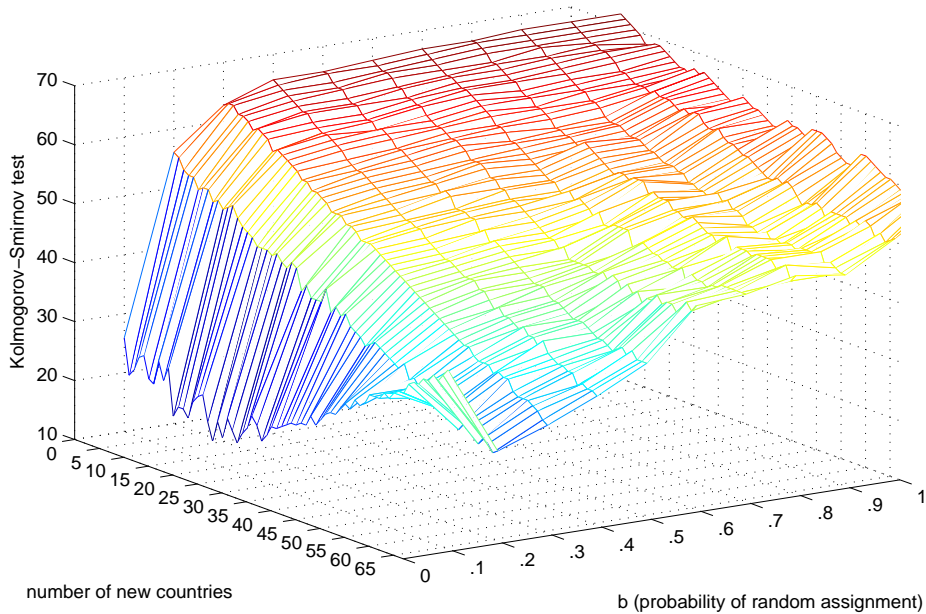}
\caption{Kolmogorov-Smirnov goodness-of-fit test for different entry rates and probabilities of random assignment}
\label{fig:KS_mesh}
\end{figure}

Figure \ref{fig:mantel} reports the value of the Mantel test for networks with $0 \leq b \leq 1$ and an entry rate $a$ which implies 
the entry form 0 to 66 countries.
The Mantel correlation statistics reach a peak of .88 ($p<.01$) in the case of preferential attachment regimes ($b=0$). 
However, the Mantel test does not discriminate among different entry regimes. 
We next compare the connectivity distribution of simulated networks with the real world distribution of 
the number of traded commodities $P(K)$ by means of the Kolmogorov-Smirnov (KS) goodness of fit test. Figure \ref{fig:KS_mesh} confirms
that the best fit is obtained in the case of a purely preferential attachment networks ($b=0$). However
the KS tests provides additional information on the most likely value of $a$ (entry rate of new nodes). 

\begin{figure}[tbhf]
\centering 
\includegraphics[width=14cm,height=9cm]{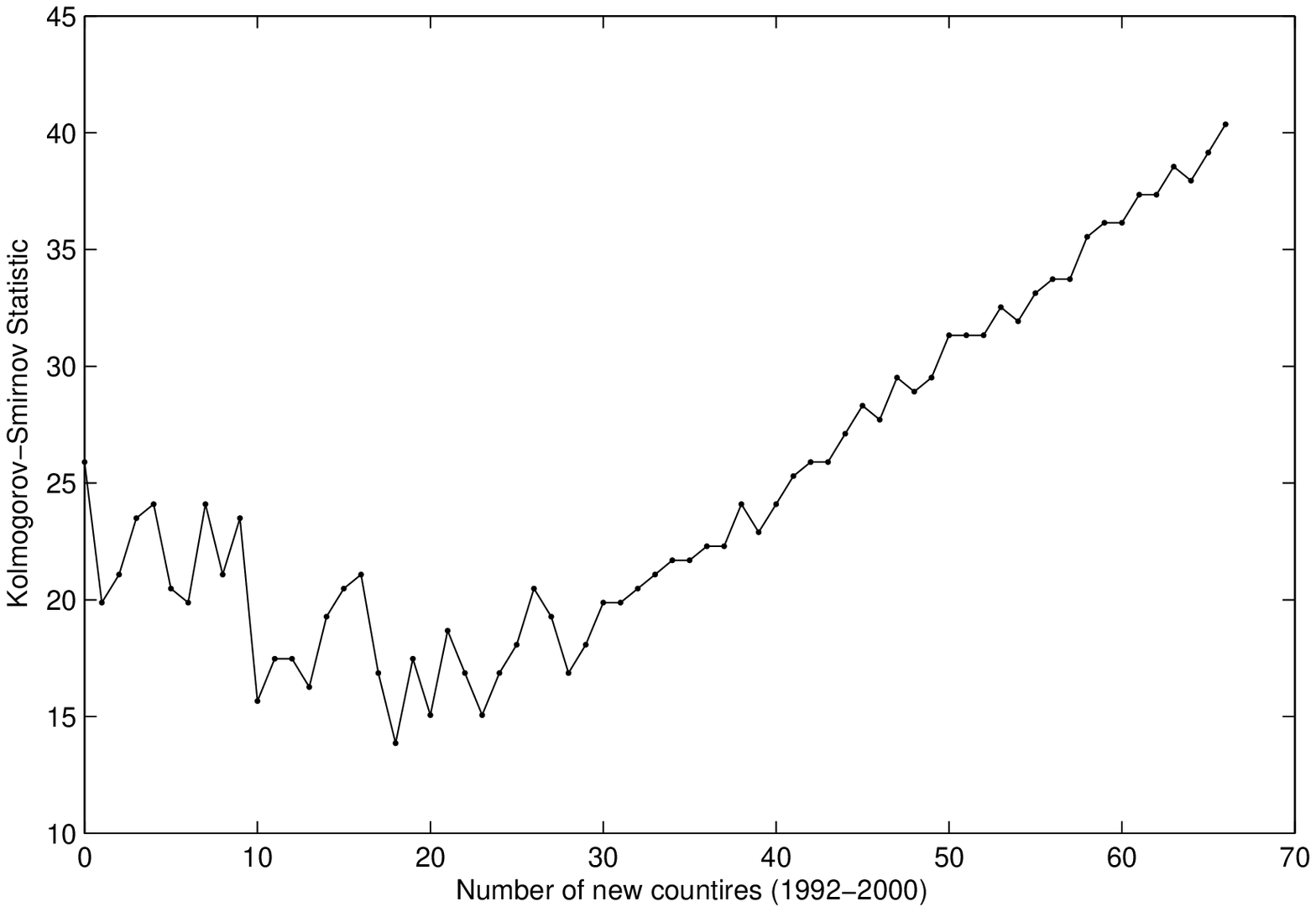}
\caption{Kolmogorov-Smirnov goodness-of-fit test for different entry rates in a pure preferential attachment regime ($b=0$)}
\label{fig:KS_test}
\end{figure}

Figure \ref{fig:KS_test} shows that the our model can better reproduce the connectivity distribution 
with and entry rate $a>0$ that implies the entry of 14--18 countries. This closely corresponds to the empirically observed number of new countries.
Thus we can conclude that a simple proportional growth model with mild entry can account for the distribution of the number of commodities traded by each pair of countries. 
 
\begin{figure}[tbhf]
\centering
\includegraphics[width = 0.9 \textwidth]{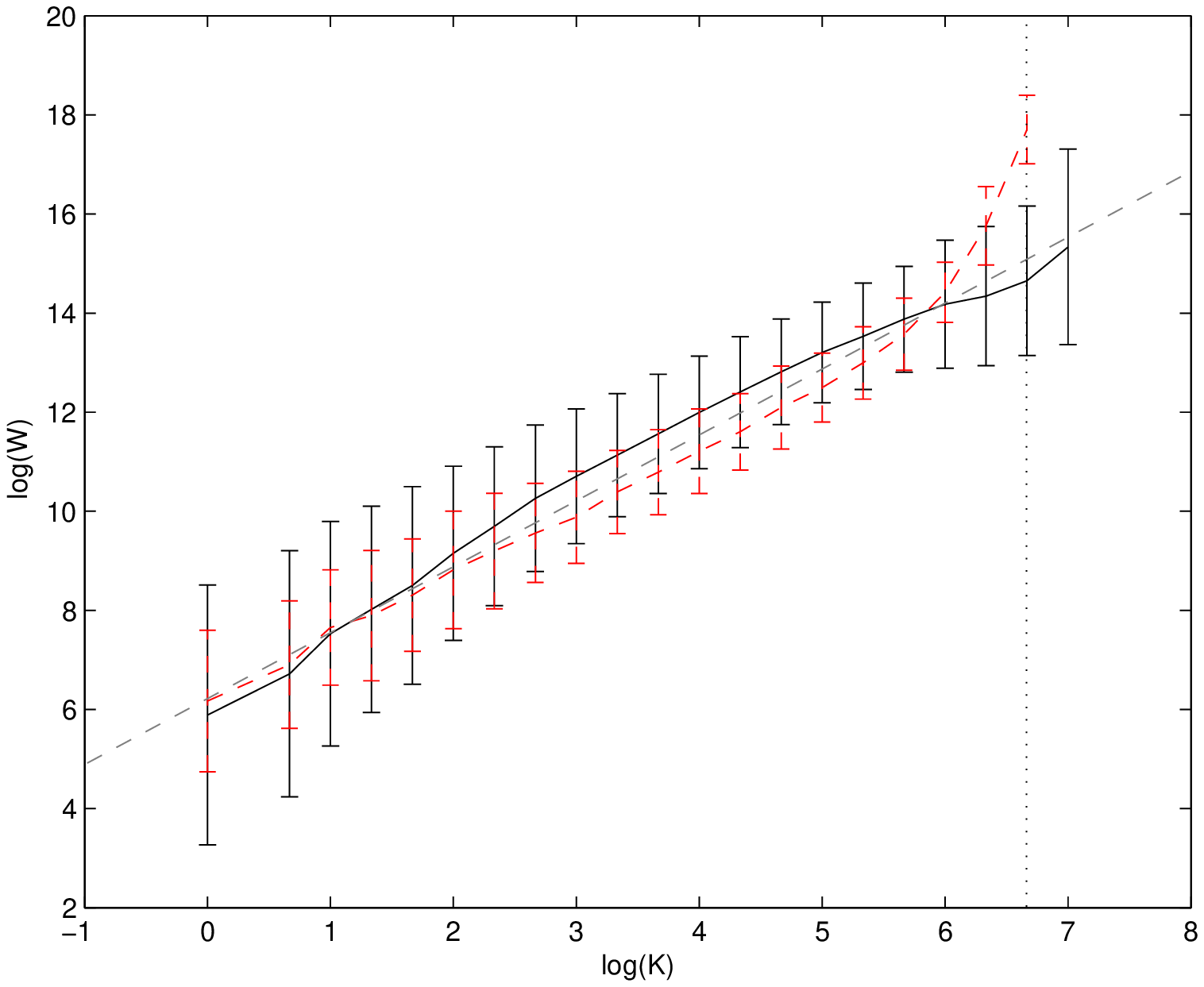}
\caption{The relationship between the number of products traded and trade value. Double logarithmic scale. Simulated (back) and real-world (red) data, mean and one standard deviation in each direction. The dashed line represents the reference line $W=K^\theta$ with $\theta\approx 1.33$}
\label{fig:size_k2}
\end{figure}

By introducing the value of the transactions we can show that the model generates the observed relationship between intensive and extensive margins of trade. 
Figure \ref{fig:size_k2} depicts the relationship between total trade flows ($W$) and the number of trade 
links maintained by each country ($K$). 
Empirically, we proxy the number of transactions by means of the number of products traded by each country.
Figure \ref{fig:size_k2} displays the relationship that emerges from 1997 trade data, and confirms that there 
exists a positive correlation between the two variables. 
The slope of the interpolating line (1.33) in double logarithmic scale reveals a positive relationship 
between the number of commodities and their average value of the kind $W=K^\theta$ with $\theta\approx 1.33$. 

The curve displays and upward departure in the upper tail. 
This can be explained by noticing that the product classification that we used imposes a ceiling to the number of products a country can trade since there are only around 1,300 4-digit categories (vertical dotted line).\footnote{Another possible explanation is that for large enough $K$, some scale effects kick in establishing a correlation between $K$ and $W$. This could be tested in other real networks with larger $K$ and no cutoff. We are aware that our modeling strategy to assume away any relationship between the mechanisms governing the binary structure of the network and the one assigning link weights is as extreme as other strategies that simply assume a single process governing the two parts. Yet, we consider the positive relationship between $K$ and $W$ as an interesting emerging property of the model.}  

Apart from the upper decile of the distribution, the simulated version of the network shows exactly the same dependence among the magnitude and the number of the transactions. 
This seems surprising by considering that the model assumes two independent growth processes for the number of transactions $K$ and their values $w$. 
However, it should be noticed that the law of large numbers does not work properly in case of skew distributions such as the log-normal. 
Given a random number of transactions with a finite expected value, if its values are repeatedly sampled from a log-normal, as the number of links increases, the average link weight will tend to approach and stay close to the expected value (the average for the population). 
However this is true only for large $K$, while according to the distribution $P(K)$ the vast majority of nodes has few links (small $K$). 
The higher is the variance of the growth process of link weights, the larger has to be $K$ to start observing  convergence toward $W= w K^\theta$ with $\theta =1$ predicted by the law of large numbers.
Thus only the largest countries approach the critical threshold.
In sum, our simulations demonstrate that our model can account for the relationship between $K$ and $W$ that has been observed in many real world weighted networks \cite{bbpsv04,eom08}.

\section{Discussion and Conclusions}
\label{sec:disc}

Using a simple model of proportionate growth and  preferential attachment we are able to replicate some of the main topological properties of real-world weighted networks.
In particular, we provide an explanation to the power-law distribution of connectivity, as well as for the fat tails displayed by the distribution of the growth rates of link weights and node strength. 
Additionally, the model matches the log-normal distribution of positive link weights 
(trade flows in the present context) and the negative relationship between node strength and variance of growth fluctuations
$\sigma(g)=W^{-\beta}$ with $\beta\approx .2$. 

The main contribution of the paper is to offer an extension of the BA model for weighted networks.
Besides, we provide further evidence that such a unifying stochastic framework is able to capture 
the dynamics of a vast array of phenomena concerning complex system dynamics \cite{fuc05}.

Further refinements of our model entail investigating its ability to match other 
topological properties of the networks such as assortativity and clustering.

\end{document}